\documentclass[10pt,english]{article}
\usepackage{mathptmx}
\usepackage[T1]{fontenc}
\usepackage[utf8]{inputenc}
\usepackage{geometry}
\usepackage{color}
\usepackage{babel}
\usepackage{float}
\usepackage{amsmath}
\usepackage{amsthm}
\usepackage{graphicx}
\usepackage{esint}
\usepackage{amsfonts}
\usepackage{amssymb}
\usepackage{lineno}
\usepackage{stackrel}
\usepackage{units}
\usepackage{mathrsfs}
\usepackage{setspace}
\usepackage[authoryear,round]{natbib} \usepackage{soul}
\usepackage{placeins}
\usepackage[unicode=true,pdfusetitle,
 bookmarks=true,bookmarksnumbered=true,bookmarksopen=true,bookmarksopenlevel=13,
 breaklinks=false,pdfborder={0 0 0},pdfborderstyle={},backref=false,colorlinks=true]
 {hyperref}

\usepackage{url} \usepackage[export]{adjustbox}[2011/08/13]

\newcommand{\citeA}[1]{\citet{#1}}
\renewcommand{\cite}[1]{\citep{#1}}

\makeatletter
\renewcommand{\@cite}[2]{#1, #2} \makeatother

\makeatletter
\newcommand{\lyxaddress}[1]{
	\par {\raggedright #1
	\vspace{1.4em}
	\noindent\par}
}

\@ifundefined{date}{}{\date{}}
\usepackage{color}
\usepackage[usenames,dvipsnames,svgnames,table]{xcolor}
\usepackage{multicol}
\definecolor{burgundy}{rgb}{0.5, 0.0, 0.13}
\definecolor{airforceblue}{rgb}{0.36, 0.54, 0.66}
\hypersetup{urlcolor=airforceblue}
\hypersetup{citecolor=burgundy}

\makeatother
\usepackage{etoolbox}
\begin{document}
\title{Dynamical constraints on the vertical structure of Jupiter's polar cyclones}
\author{\href{https://orcid.org/0000-0002-3645-0383}{Nimrod Gavriel}$^{1\star}$
and \href{https://orcid.org/0000-0003-4089-0020}{Yohai Kaspi}$^{1}$}
\maketitle

\lyxaddress{\begin{center}
\textit{$^{1}$Department of Earth and Planetary Sciences, Weizmann
Institute of Science, Rehovot, Israel}\\
\textit{$^{\star}$\href{mailto:nimrod.gavriel@weizmann.ac.il}{nimrod.gavriel@weizmann.ac.il}}
\par\end{center}}

\lyxaddress{\begin{center}
Preprint November 2, 2025\textit{}\\
This article has been published in \\
\textit{Proceedings of the National Academy of Sciences (PNAS)} (2025), 122 (44) e2503737122. \\
The final version is available at DOI:\href{https://doi.org/10.1073/pnas.2503737122}{10.1073/pnas.2503737122}. 
\\
(Received Feb 19, 2025; Revised June 24, 2025; Accepted September 16, 2025) 
\par\end{center}}

\begin{abstract}
Jupiter's poles feature striking polygons of cyclones that drift westward over time, a motion governed by $\beta$-drift (vortex motion caused by the latitudinal variation of the Coriolis force). This study investigates how $\beta$-drift and the resulting westward motion depend on the depth of these cyclones. Counterintuitively, shallower cyclones drift more slowly, a consequence of stronger vortex stretching. By employing a 2D quasi-geostrophic model of Jupiter's polar regions, we constrain the cyclones' deformation radius, a key parameter that serves as a proxy for their vertical extent, required to replicate the observed westward drift. We then explore possible vertical structures and the static stability of the poles by solving the eigenvalue problem that links the 2D model to a 3D framework, matching the constrained deformation radius. These findings provide a foundation for interpreting upcoming Juno microwave measurements of Jupiter's north pole, offering insights into the static stability and vertical structure of the polar cyclones. Thus, by leveraging long-term motion as a novel constraint on vertical dynamics, this work sets the stage for advancing our understanding of the formation and evolution of Jupiter's enigmatic polar cyclones.
\end{abstract}

\twocolumn 

\subsection*{Key Points}
\begin{itemize}
\item $\beta$-drift of cyclones, linked to Jupiter's polar cyclones' westward drift, strengthens for deeper cyclones as vortex stretching diminishes.
\item A 2D quasi-geostrophic model estimates a deformation radius of about 220 km for the north pole and 360 km for the south pole.
\item The estimated deformation radii uncover possible vertical structures shaped by vertical stability. \end{itemize}

\section*{Plain language summary} 
Jupiter’s poles feature clusters of large, long-lasting cyclones that slowly drift westward. We argue that the depth to which these cyclones extend below the cloud layers significantly influences this drift rate. By comparing computer simulations with observations, we estimate the depth of the cyclones and shed light on the stability of Jupiter’s atmosphere at high latitudes. These insights are particularly critical for Juno’s extended mission, which now allows the Microwave Radiometer (MWR) to observe these cyclones with unprecedented resolution, providing a direct test of our framework and facilitating quantitative interpretation of their vertical structure. Our results directly support this new phase of the Juno mission, offering key constraints to understand the formation, persistence, and dynamics of these giant polar storms.
 
\section{Introduction}
\begin{figure*}[t!]
\includegraphics[width=1.03\textwidth,center]{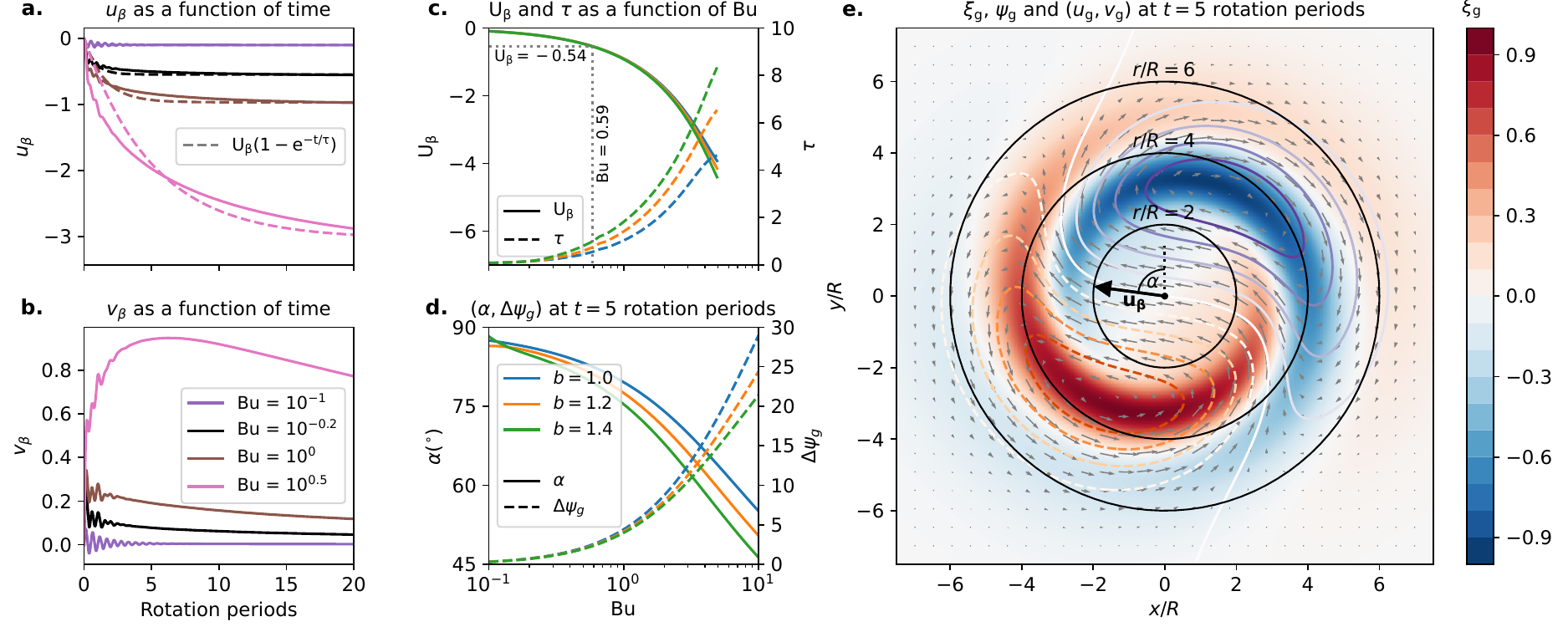}
\caption{\textbf{Impact of Stretching on $\beta$-Drift Dynamics in a Single-Layer QG Framework: A Single Cyclone on a $\beta$-Plane.}
(a--b) Evolution of the zonal ($u_\beta$) and meridional ($v_\beta$) drift velocities over time for four different values of the Burger number and $b=1.2$. Dashed curves show exponential fits with amplitude $\rm{U}_\beta$ and timescale $\tau$. Here, $(u_\beta,v_\beta)$ are non-dimensional, scaled by $\hat{\beta}V$.
(c) Plots of $\rm{U}_\beta$ (solid line, left axis) and $\tau$ (dashed line, right axis, in rotation periods) as functions of ${\rm Bu}$ for three values of the shape factor $b$ (legend in panel~d). The exponential parameterization becomes inadequate for ${\rm Bu} \gtrsim 5$ (see Fig.~S2).
(d) Solid line: Phase angle $\alpha$ (defined in panel~e) between $u_\beta$ and $v_\beta$ as a function of  ${\rm Bu}$. Dashed line: Strength of the $\beta$-gyres, quantified as $\Delta\psi_g=\max(\psi_g)-\min(\psi_g)$, with values non-dimensionalized by $\hat{\beta}RV$.  
(e) The vorticity field ($\xi_{\rm g}$, in color), streamfunction ($\psi_{\rm g}$, in contours), and velocity field ($\mathbf{u}_{\rm g}$, arrows) after five rotation periods for $\rm{Bu}=0.59$ and $b=1.2$. The $\beta$-gyres appear as red and blue lobes surrounding the cyclone core. The colorbar represents the scaled (by $\hat{\beta}V/R$) vorticity field. See Movie~S1 for the simulation.
 }
\label{fig: beta-drift with Bu}
\end{figure*}
\subsection{Probing beneath the clouds of Jupiter}

Historically, our understanding of Jupiter's atmosphere has been limited to observations of its visible cloud layer, located near the 1-bar pressure level. This limitation arises from the challenges of probing beneath the optically thick ammonia and water clouds \cite{atreya1999, Pater2016}. Until recently, the only direct measurements of Jupiter's subcloud layers were the localized and costly in-situ data collected by the Galileo Probe \cite{young1998galileo}. However, an indirect method using gravity measurements (by tracking changes in the Juno spacecraft's acceleration during close flybys) has enabled insights into the large-scale flows at depth. This approach has revealed the 3D structure of zonal flows within Jupiter's subequatorial jet streams \cite{kaspi2018jupiter, galanti2021, Kaspi2023} and constrained the depth of the Great Red Spot (GRS) to less than $500$~km \cite{galanti2019determining, Parisi2021}.

Another approach to exploring Jupiter's subcloud layers involves observations at cloud-penetrating frequencies. For example, using radio frequencies, the Earth-based Very Large Array (VLA) telescopes have probed down to pressures of approximately 8 bars \cite{Pater2016}. The Juno Microwave Radiometer (MWR) offers an enhanced use of cloud-penetrating frequencies, leveraging its orbital proximity to achieve deeper measurements with finer spatial resolution. Equipped with six channels, the MWR has been providing spatially resolved observations of brightness temperature down to depths of approximately 240 bars \cite{bolton2017jupiter, Janssen_2017}. These thermodynamic measurements are influenced both by temperature variations \cite{Fletcher2021} and the opacity of ammonia clouds \cite{Li2017ammonia}. Although there is ongoing debate about the relative contributions of these factors, the observed deep patterns reveal valuable information about the vertical structure and depth of atmospheric phenomena seen at the cloud tops. For example, the MWR detected alternating zonal bands of ammonia depletion and enrichment extending to 240 bars between midlatitude jets. These patterns correspond to the upwelling and downwelling branches expected in a Ferrel-cell-like circulation \cite{Duer2021, Duer2023}.

Another key instrument on Juno, the Jovian Infrared Auroral Mapper (JIRAM), was designed to study the aurora and atmospheric chemistry at 5-7 bar \cite{Adriani2017}. During Juno's polar orbits, as JIRAM is insensitive to half the pole being dark and can wholly observe the pole, it discovered a new phenomenon: a polar "crystal" of massive cyclones (each $\sim5,000$ km wide) inhabiting each pole, consisting of a polar cyclone (PC) and 8 circumpolar cyclones (CPCs) at the north pole, and 5 at the south \cite{adriani2018}. The MWR instrument, designed to function optimally at the shorter distances from Jupiter's cloud-tops during the around-equatorial skims in its original polar orbit, now enters a phase in Juno's extended mission, where the orbit has considerably shifted, allowing MWR measurements in near north pole of Jupiter \cite{Orton2024epsc}. This enables multi-channel measurements with sufficient resolution to capture the north polar cyclones, similar to analogous previous measurements of vortices in the midlatitudes \cite{Bolton2021}. This study aims to provide a theoretical framework for interpreting the upcoming MWR observations of Jupiter’s north pole, offering predictions about the vertical structure and extent of the polar cyclones.

\subsection*{Jupiter's Polar Cyclones}

The energetic processes sustaining Jupiter's polar cyclone configurations, while explored in prior studies \cite{oneill2015polar,  Brueshaber2019, hyder2022, siegelman2022moist, Siegelman2022}, remain largely unconstrained. Nevertheless, substantial progress has been made in understanding the principal momentum balance that governs these cyclones, explaining both their stability and long-term motion. This momentum balance is closely related to the concept of $\beta$-drift, a secondary effect that causes vortices to move due to their interaction with a background vorticity gradient \cite{sutyrin1994intense} known to influence the motion of tropical cyclones on Earth \cite{franklin1996tropical}. The phenomenon of 
$\beta$-drift has been explored extensively in idealized settings, including theoretical studies \cite{adem1956series, smith1990analytical}, numerical simulations \citeA{lam2001, Gavriel2023}, and laboratory experiments \cite{Firing1976,Carnevale1991,STEGNER1998,FLOR2002,Benzeggouta2025}.

Traditionally, as suggested by the term "$\beta$-drift", studies have focused on the interaction of cyclones with $\beta$, the meridional gradient of the planetary vorticity $f$, causing a poleward-westward migration through the generation of opposite vorticity anomalies (usually termed $\beta$-gyres) surrounding the core. However, in addition to $\beta$, cyclones on Jupiter's poles exhibit mutual repulsion due to the interaction of the vorticity gradient of one cyclone with the velocity field of another. This mutual interaction allows for the stable crystal-like configuration of the cyclones \cite{Gavriel2021}, provided that they are surrounded by an anticyclonic annulus, commonly referred to as shielding \cite{li2020}. This repulsion also explains the oscillatory motion of the polar cyclones \cite{Gavriel2022}, as observed by Juno \cite{mura2021oscillations}.

Beyond oscillatory motion, the cyclones exhibit a mean westward drift, with rates of approximately $3^\circ$ and $7^\circ$ per year at the north and south poles, respectively \cite{adriani2020two, mura2022five_years}. This westward drift arises from the $\beta$-drift acting collectively on the group of cyclones. By adopting a "center of mass" perspective, the mutual interactions between the cyclones are effectively averaged out, leaving only the cumulative interaction of the group with $\beta$. This dynamic leads to a group oscillation around the poles accompanied by a westward precession, manifesting as the observed drift \cite{Gavriel2023}.

In this study, we investigate how the vertical structure and depth of Jupiter's polar cyclones influence their mean westward drift. First, we employ an idealized 2D model to explore how depth, and the associated column stretching, affects $\beta$-drift on a vortex in a $\beta$-plane. Next, we perform a parameter scan using a 2D Quasi-Geostrophic (QG) polar model with Jupiter's polar cyclones initialized in the system to constrain the QG deformation radius ($L_d$), a proxy for depth, such that the modeled westward drift matches observational data. Finally, we solve the eigenvalue problem posed by comparing 2D and 3D QG systems to deduce the vertical structures and static stability of the polar cyclones based on the constrained $L_d$.

\begin{figure*}[t!]
\includegraphics[width=0.92\textwidth,center]{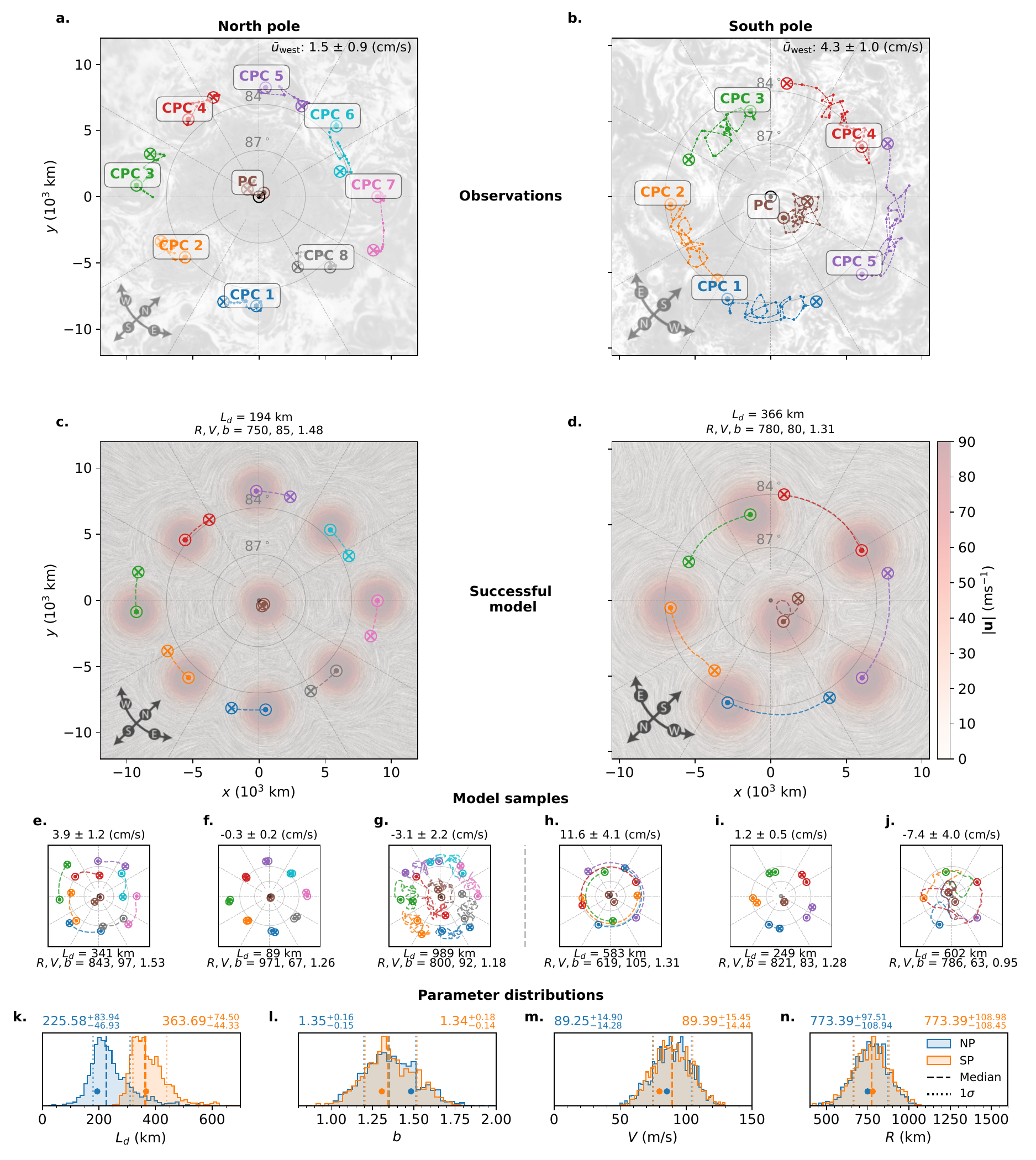}
\caption{\textbf{Constraining $L_d$ at Jupiter's Poles Using Observed and Simulated Cyclone Drifts}. 
(a,b) Observed trajectories of the north (a) and south (b) polar cyclones over five years \cite{mura2022five_years}. The symbols $\odot$ and $\otimes$ mark the start and end points, respectively. The background is a JIRAM infrared image from PJ4 (Image from \citeA{adriani2018}), corresponding to the onset of these trajectories. 
(c,d) Simulated trajectories after five years for a model that reproduces the observed mean westward drift. The color scale indicates the magnitude of the velocity vector, while the streaks represent the flow direction through line integral convolution (LIC) of the velocity field at $t=0$. 
(e--j) Sample model trajectories from the nested sampling routine, illustrating diverse outcomes. The numbers above the panels are the mean and standard deviation (between the cyclones) of the westward drift. The parameters used in each panel are listed below the panels.
(k--n) Histogram plots of the posterior parameter distributions produced by Ultranest \cite{buchner2021ultranest}, after sampling $15{,}000$ parameter sets per pole for the north (blue) and south (orange) polar cyclones. The dashed lines denote the median, and the dotted lines mark $\pm1\sigma$ for the distributions. These values are written above the panels. Markers represent the parameter values of the successful models (panels c,d). For a corner plot of the joint posterior distributions, see Fig.~S4.
 }
\label{fig: 2D polar model fig depth}
\end{figure*}
\FloatBarrier
\section*{The Impact of Vertical Depth on Vortex $\beta$-Drift}

A standard framework for understanding the motion of vortices due to $\beta$-drift is the barotropic QG equation of a seeded vortex on a $\beta$-plane \cite{smith1990analytical, sutyrin1994intense, smith1991analytic}. In this formulation, cyclones undergo poleward-westward propagation in a planetary $\beta$, with the drift amplitude scaling with the dimensionless parameter $\hat{\beta}=\frac{\beta R^2}{V}$ \cite{Smith_Llewellyn1997,Gavriel2023}, where $R$ and $V$ are the characteristic size and velocity of the cyclone. In this section, we seek to understand how the vertical extent (through stretching) modifies the classical $\beta$-drift by adding the stretching term to the QG equation, before treating other important complexities such as vortex-vortex interactions.  Stretching, in this context, refers to the vertical elongation (or compression) of a fluid column accompanied by horizontal convergence (or divergence). In QG dynamics, such vertical motions alter the potential vorticity (PV) by concentrating or diluting the background vorticity (see Fig.~S1 and the accompanying text for more intuition). 

Following the derivation in \citeA{smith1990analytical} but including stretching (see SI for scaling, derivation, and other details), we develop the equations in a reference frame moving with the center of the cyclone. The scaled resulting QG equation becomes (focusing on the dominant terms)
\begin{equation}
\frac{\partial}{\partial t}\left(\xi_{\rm g} - \frac{\psi_{\rm g}}{\rm Bu}\right)
= -\hat{\beta}v_{\rm v} - \mathbf{u}_{\rm v}\cdot\mathbf{\nabla}\xi_{\rm g},
\label{eq: QG for vortex drift}
\end{equation}
where $\xi$ and $\psi$ are the vorticity and streamfunction, respectively, and $\mathbf{u}=(u,v)$ is the velocity vector. Subscripts $\rm v$ and $\rm g$ represent, respectively, a constant background vortex flow and a "generated" field leading to the $\beta$-drift. The Burger number, ${\rm Bu} \equiv \left(\frac{L_d}{R}\right)^2$, incorporates the effect of the deformation radius, $L_d = \frac{\sqrt{gH}}{f_0}$, where $g$ is gravitational acceleration, $H$ is the layer depth, and $f_0$ is the Coriolis parameter. The term $-\frac{\partial}{\partial t}\left(\frac{\psi_{\rm g}}{\rm Bu}\right)$ captures the influence of stretching in the single-layer QG model.

The model is solved for vortex profiles of the dimensional form \cite{chan1987analytical}:
\begin{equation}
\xi_{\rm v} 
= \frac{V}{R} \left( 2 - \left(\frac{r_{\rm c}}{R}\right)^b \right) 
\exp\left[\frac{1}{b}\left(1 - \left(\frac{r_{\rm c}}{R}\right)^b\right)\right],
\label{eq: chan87 vorticity profile}
\end{equation}
where $r_{\rm c}$ is the distance from the vortex's center, and $b$ is a shape factor that controls the sharpness and shielding of the vortex profile. For a more detailed analysis of the role of $b$, see \cite{li2020}. The initial-value problem is solved using the Dedalus PDE solver \cite{burns2020dedalus}. The results from this model, as a function of ${\rm Bu}$, are shown in Fig.~\ref{fig: beta-drift with Bu}. The zonal $\beta$-drift velocity, defined as $u_\beta \equiv u_{\rm g}{\mid_{(x=0, y=0)}}$, displays an exponential relaxation with time, which is modeled by $\rm{U}_\beta\bigl(1 - e^{-t/\tau}\bigr)$ (Fig.~\ref{fig: beta-drift with Bu}a), a parameterization that provides a good fit for $\rm{Bu}\lesssim
5$. 

We further examine the dependence of the fitted maximum drift velocity ($\rm{U}_\beta$) and timescale ($\tau$) on layer depth (Fig.~\ref{fig: beta-drift with Bu}c) for three values of the shape factor $b$ within a physically relevant range. The results indicate that while the drift amplitude is relatively insensitive to changes in $b$, it becomes more negative (more strongly westward) as the Burger number increases (or equivalently, as $H$ increases).

In \textit{Gavriel and Kaspi, 2023} \cite{Gavriel2023} (their Fig.~2), the layer depth $H$ is assumed to be infinite, which implies $L_d \rightarrow \infty$ and ${\rm Bu} \rightarrow \infty$, effectively removing the stretching term from Eq.~\ref{eq: QG for vortex drift}. This assumption highlights that for very deep layers, any change in the fluid upper surface height $(\delta h)$ is negligible compared to the total depth $(\frac{\delta h}{H}\ll1)$, rendering stretching effects unimportant to the dynamics.

However, when $H$ (and thus ${\rm Bu}$) is finite, the PV anomaly generated by the $\beta$-effect (the term $-\hat{\beta} v_{\rm{v}}$) is divided between two parts: the formation of $\beta$-gyres (through $\frac{\partial \xi_{\rm g}}{\partial t}$) and column stretching (through $-\frac{\partial}{\partial t}\frac{\psi_{\rm g}}{\rm Bu}$). As stretching has more relative significance in shallower layers, it claims a larger share of the total PV response, thereby weakening the $\beta$-gyres.
This trend is reflected in Fig.~\ref{fig: beta-drift with Bu}c, where shallower layers correlate with stronger stretching and hence smaller $\beta$-drift velocities. This reduction is also evident in Fig.~\ref{fig: beta-drift with Bu}d, where the difference between the extreme streamfunction values of the $\beta$-gyres, $\Delta\psi_g$, becomes smaller.

Using the Jovian values for the north and south poles of $\hat{\beta}$, $R$, and $V$ from \citeA{Gavriel2023}, we find that matching the observed westward drift corresponds to $\rm{U}_\beta=-0.46$ for the north pole and $\rm{U}_\beta=-0.54$ for the south pole, marked in Fig.~\ref{fig: beta-drift with Bu}c. These values correspond to ${\rm Bu}=0.49$ for the north pole and ${\rm Bu}=0.59$ for the south, which is used in Fig.~\ref{fig: beta-drift with Bu}e to show the generated fields after five rotation periods. The influence of ${\rm Bu}$ extends beyond the reduction of the $\beta$-drift amplitude. Smaller ${\rm Bu}$ values also reduce the time required to reach the peak drift velocity ($\tau$ in Fig.~\ref{fig: beta-drift with Bu}c) and lead to a more zonally directed drift overall ($\alpha$ in Fig.~\ref{fig: beta-drift with Bu}d).

From ${\rm Bu}=(0.49,0.59)$ and $R=(811,861)$\,km \cite{Gavriel2023}, the resulting deformation radii are $L_d=(568,658)$\,km for the north and south poles, respectively.  These values should not be interpreted as definitive physical estimates, but rather as outcomes of an idealized model that omits vortex–vortex interactions and the polar variation of $f$. Instead, this simplified framework is meant to provide insight into the role of stretching alone in modulating $\beta$-drift. In the next section, we consider these additional complexities to derive more realistic values of $L_d$ needed to match the observed drift rates.

 \section*{Estimating the Deformation Radius Using a Single-Layer QG Model of the Jovian Poles}

While the previous section focused on how vertical extent modifies $\beta$-drift, the idealized model used there does not provide a realistic estimate of $L_d$. To address this limitation, we now present a single-layer QG model initialized with a polar vortex crystal. This model evolves under the full spherical form of the vertical component of Jupiter's rotation rate, $f = 2\Omega \cos\left(\frac{r}{R_{\rm J}}\right)$,
where $\Omega$ is Jupiter's rotation rate, $R_{\rm J}$ is Jupiter's radius, and $r$ is the distance from the pole. We use Dedalus \cite{burns2020dedalus} to solve the polar QG equation \cite{vallis2017atmospheric}:
\begin{equation}
\frac{D}{Dt}
\left(\nabla^2\psi 
+ 2\Omega \cos \left(\frac{r}{R_{\rm J}}\right)
- \frac{1}{L_{d}^2}\psi 
\right) = 0,
\label{eq: 2D QG equation, depth}
\end{equation}
with multiple cyclones prescribed by Eq.~\ref{eq: chan87 vorticity profile}, with one PC and eight (five) CPCs placed initially at the observed locations during Juno's PJ4 orbit for the north (south) poles (Fig.~\ref{fig: 2D polar model fig depth}a-b). For simplicity and because of limited observational data, we assume identical cyclones. Due to this choice and the absence of small-scale forcing, the oscillatory motion discussed by \citeA{Gavriel2022} is not excited. Hence, we focus on reproducing the mean westward drift. While zonal jets could, in principle, influence the drift, no such flows have been observed in the vicinity of the cyclones \cite{Rogers2022}. Moreover, JunoCam imagery is consistent with a two-dimensional, isotropic turbulent regime \cite{orton2017first}, and thus zonal flows are not included in this model.

This model involves four unknown parameters: $L_d$, and the three cyclone profile parameters $(R, V, b)$. To explore the phase space of this model, we use the Ultranest library \cite{buchner2021ultranest}, which employs a Bayesian nested sampling approach to identify the best-fitting solutions \cite{buchner2023nested}. As a benchmark, we use the 5-year observed trajectories of the north and south polar cyclones \cite{mura2022five_years} (Fig.~\ref{fig: 2D polar model fig depth}a-b) to calculate the average westward drift between the cyclones. This value is then used to calculate a likelihood function that evaluates the probability of a parameter set being accurate, based on the resultant model westward drift (see SI for details on the numerical model, likelihood function, and model sampling).

We sampled $15{,}000$ parameter sets for each pole, integrating each simulation for seven years and calculating the resultant mean westward drift between the cyclones during the last five years (where the first 2 years act as a spin-up). Illustrative model outputs (Fig.~\ref{fig: 2D polar model fig depth} panels e--g for the north pole and panels h-j for the south) span a variety of outcomes: drifts too fast (panels~e,h), too slow (panels~f and i), eastward (panel~g), and mergers (panel~j). Examples of parameter sets that successfully recreate the observed drift are shown in Fig.~\ref{fig: 2D polar model fig depth}c-d.

Due to intrinsic parameter degeneracies, the algorithm returns a posterior distribution rather than a specific set of parameters that matches the observed westward drift. From this distribution, we obtain a more realistic estimate for the deformation radius of $L_d = 225.58_{-46.93}^{+83.94}$~km and $L_d = 363.69_{-44.33}^{+74.50}$~km at the $1\sigma$ confidence level for the north and south poles (Fig.~\ref{fig: 2D polar model fig depth}k), respectively. The resulting posterior distributions for $R$, $V$, and $b$ (Fig.~\ref{fig: 2D polar model fig depth}l--n) align well with plausible mean values for Jupiter's polar cyclones, as estimated from cloud feature tracking at the poles \cite{grassi2018first}.

 \begin{figure*}[ht!]
\includegraphics[width=1.15\textwidth,center]{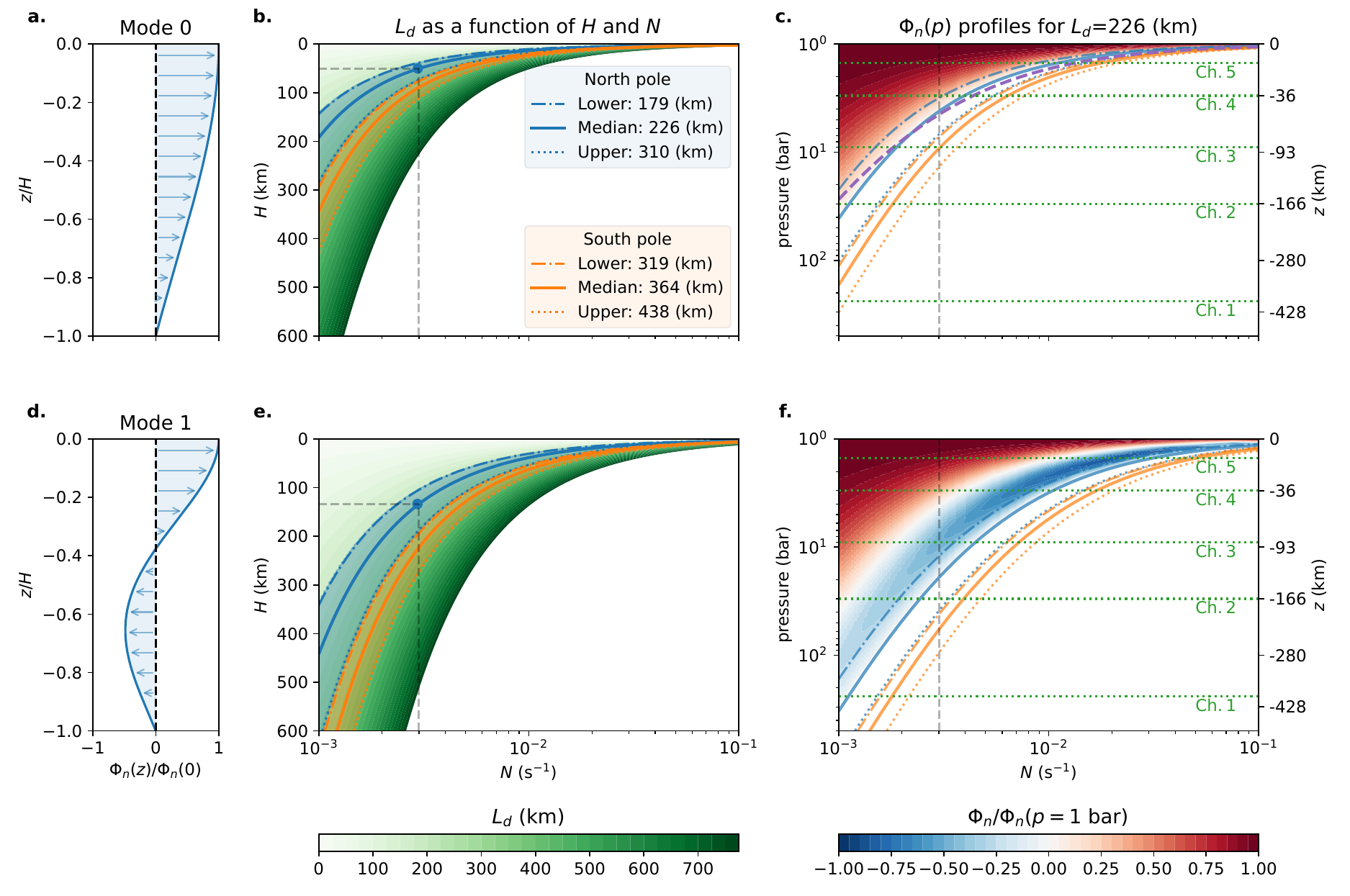}
\caption{\textbf{Interpretation of the Cyclones' Vertical Structure from the Deformation Radius}. The first row corresponds to the "mode~0" solutions of the eigenvalue problem described by Eq.~\ref{eq: eigenvalue problem, depth}, while the second row corresponds to "mode~1". (a,d) Vertical profiles of $\Phi_n$ for each mode. (b,e) Solutions for $L_d$ (green shades) for various combinations of $N$ (Brunt-V\"ais\"al\"a frequency) and $H$ (depth). The estimated $L_d$ distributions from Fig.~\ref{fig: 2D polar model fig depth}k are shown by the blue and orange curves for the north and south poles, respectively. (c,f) Solutions for $\Phi_n$ as a function of pressure below the cloud layer, following the $N$ and $H$ relationship indicated by the blue solid curve in panels (b) and (e). Green horizontal dotted lines indicate the depths at which each MWR channel has maximum sensitivity \citep{Janssen_2017}, though each channel is sensitive over a broader range of depths. The numbers on the right ordinate show the altitude relative to 1~bar. Grey dashed lines in panels (b,c,e,f) denote the values used to generate the profiles in panels (a) and (d), and to produce Fig.~\ref{fig: 3D fig}. The color along the grey dashed line in panels (c,f) corresponds to the $\Phi(z)$ profiles shown in panels (a,d). The purple dashed line in panel (c) represents the theoretical scaling $H \sim \mathrm{Ro}^{1/2} R f_0/N$ (adapted from \citep{Aubert2012}), using representative values of $R=800 \,\rm{km}$ and $U=90\,\rm{ms}^{-1}$ from  Fig.~\ref{fig: 2D polar model fig depth}m-n.
 }
\label{fig: Depth eigenprob}
\end{figure*} 

\section*{Interpreting the Vertical Structure of Jupiter's Polar Atmosphere from the Drift Rate of the Polar Cyclones}

The standard shallow-water (SW) expression for the deformation radius, $L_d = \sqrt{gH}/f_0$, is not well-suited to Jupiter's upper atmosphere. This expression assumes a constant density and no vertical dependence of the flow, whereas Jupiter's upper-atmospheric density increases exponentially with depth \cite{Guillot1995b} over many scale heights. To account for these vertical variations, we consider the continuously stratified QG vorticity equation \cite{vallis2017atmospheric}:
\begin{equation}
\frac{D}{Dt}\left(\nabla^{2}\psi+f+\frac{f_{0}^{2}}{\tilde{\rho}}\frac{\partial}{\partial z}\left(\frac{\tilde{\rho}}{N^{2}}\frac{\partial\psi}{\partial z}\right)\right)=0,
\label{eq: 3D QG equation, depth}
\end{equation}
where $N$ is the Brunt-V\"ais\"al\"a  frequency, indicating the stability of the atmosphere with respect to vertical motion, and $\tilde{\rho}$ is the vertical density profile. 

To account for variations in the vertical direction, we assume the streamfunction can be separated into horizontal and vertical components:
\begin{equation}
\psi(x,y,z,t)=\sum_n \psi_n(x,y,t)\Phi_n(z)
\end{equation}
where $\Phi_n(z)$ is the vertical profile for each $n$. Substituting this form into the QG equations requires each $\Phi_n$ to satisfy the vertical eigenvalue problem \cite{Smith2001b}:
\begin{equation}
\frac{f_{0}^{2}}{\tilde{\rho}}\frac{\partial}{\partial z}\left(\frac{\tilde{\rho}}{N^{2}}\frac{\partial\Phi_n}{\partial z}\right)+\Gamma_n\Phi_n=0,
\label{eq: eigenvalue problem, depth}
\end{equation}
where $\Gamma_n$ is an eigenvalue. For each mode $n$, the corresponding deformation radius is given by $L_{d,(n)}=\Gamma_n^{-\frac{1}{2}}$ (see SI for the full derivation). Assuming that the cyclones are dominated by a single vertical mode, the horizontal dynamics derived from the two-dimensional QG equation (Eq.~\ref{eq: 2D QG equation, depth}), and the three-dimensional QG equation (Eq.~\ref{eq: 3D QG equation, depth}) will yield similar results for the given $L_{d,(n)}$; the linearized dynamics are identical (see SI). Given our estimates of the $L_d$ values required to achieve the observed westward drift (Fig. \ref{fig: 2D polar model fig depth}), we can use the eigenvalue problem posed by Eq.~\ref{eq: eigenvalue problem, depth} to investigate which vertical structures (eigenfunctions $\Phi_n(z)$) satisfy the required $L_d$.

We solve this eigenvalue problem using the Dedalus library \cite{burns2020dedalus}, applying boundary conditions of no vertical velocity at the tropopause $(\partial_z\Phi_n|_{_{z=0}}=0)$ and no flow at the bottom $(\Phi_n|_{_{z=-H}}=0)$. For $\tilde{\rho}(z)$, we adopt a reference density profile derived from interior models of Jupiter, constrained by a neural-network-based parameter sweep \cite{Ziv2024,Ziv2024a}. We focus here on the two lowest modes ($n=\{0,1\}$), which are illustrated in Fig.~\ref{fig: Depth eigenprob}(a,d). The $n=0$ mode corresponds to a simple decay profile, whereas $n=1$ changes sign along the column. This focus on the two leading modes aligns with estimates from Earth's oceans, where they capture the bulk of the kinetic energy \cite{Smith2001b,Ferrari2010}; higher modes require excitations with elaborate vertical profiles.

Since Eq.~\ref{eq: eigenvalue problem, depth} depends only on $H$ and $N$ (here assumed constant, see discussion in the SI), we present $L_{d,(n)}$ solutions as a function of these two parameters in Fig.~\ref{fig: Depth eigenprob}(b,e), for both modes. The constrained range of deformation radii from the previous section  (Fig.~\ref{fig: 2D polar model fig depth}k) is indicated for the north (blue) and south (orange) poles, essentially giving an $H$ as a function of $N$ for each pole. Focusing on the relationship defined by the blue solid curve in Fig.~\ref{fig: Depth eigenprob}(b,e), we plot the corresponding $\Phi_n(z)$ profiles as a function of pressure below the cloud level for different $N$ values (Fig.~\ref{fig: Depth eigenprob}(c,f)). 

Interestingly, in agreement with the eigenvalue results, the $H$ vs $N$ line in Fig.~\ref{fig: Depth eigenprob}c closely follows a scaling law derived for the aspect ratio of vortices in stratified fluids, supported by theory and rotating tank experiments \citep{Billant2001, Aubert2012, lemasquerier2020remote}. This vertical scaling, $H \sim f_0 R\sqrt{\mathrm{Ro}\left(1-\mathrm{Ro}\right)/\left(N_c^2 - N^2\right)} $, can be simplified in our case to $H \sim  f_0 R\,\sqrt{\mathrm{Ro}} /N$. This approximation is valid under the assumptions that the Rossby number is small ($\mathrm{Ro} = U/(f_0 R) \ll 1$) and that the static stability at the vortex center is small ($N_c^2\ll N^2$). Under these conditions, the simplified scaling (purple dashed line in Fig.~\ref{fig: Depth eigenprob}c) provides a good match to the curve derived from the eigenvalue solutions for the North Pole.

The MWR instrument, expected to probe below Jupiter's clouds at the north pole in the near future, measures brightness temperature at six channels. The maximum sensitivity pressure levels for each MWR channel are marked by green dotted lines, with Channel~6 above the 1-bar level. For a plausible range $N=3 - 20 \times 10^{-3}  \rm{s}^{-1}$ \cite{Lee2021}, consistent with our reanalysis of Galileo probe data (see SI and Fig.~S5), and focusing first on mode~0 (Fig. \ref{fig: Depth eigenprob}c), we anticipate that the north polar cyclones' footprint will be visible in Channels~6 and 5 (and Channel 4 if $N\approx 3 \times 10^{-3}  \rm{s}^{-1}$). However, if the real Jovian poles exhibit turbulence with a vertical phase shift, reminiscent of terrestrial storms with low-level convergence and upper-level divergence, then mode 1 may also be excited. In that scenario (Fig.~\ref{fig: Depth eigenprob}f), signatures could extend into Channel~3 for the same range of $N$s.

\begin{figure}[t!]
\includegraphics[width=0.57\textwidth,center]{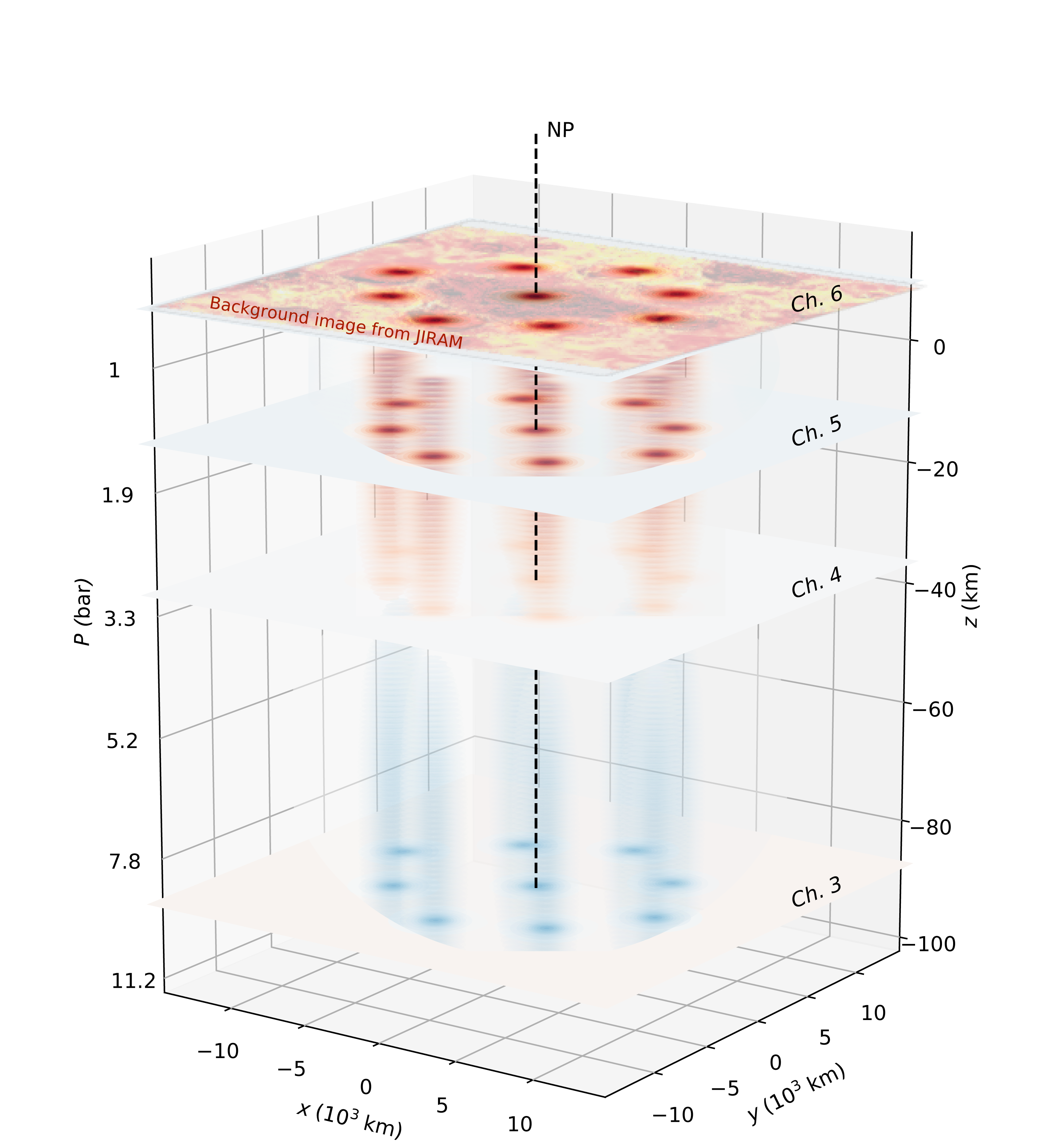}
\caption{\textbf{Vertical Structure and Intersections with Instrument Sensitivity Depths at the North Pole}. The $\psi$ field from Fig.~\ref{fig: 2D polar model fig depth}c was extended downward (rather than computed directly in a 3D simulation) using the eigenfunctions from Fig.~\ref{fig: Depth eigenprob}a,d, which correspond to $N = 3 \times 10^{-3}\;(\mathrm{s}^{-1})$ and $H = 50$ ($133$) km for mode 0 (mode 1). Both modes yield $L_d = 226$ km, with mode $0$ terminating ($\psi=0$) at $z=-50$ km, and mode 1 continuing further to $\psi=0$ at $z=-134$ km. The planes represent $\psi$ at the MWR channels that overlap with the eigenfunction range, while the shaded regions between the planes depict $\psi$ values between the channels. The dashed black line indicates the north pole. The background image on channel 6 is a JIRAM measurement from PJ4 \cite{adriani2018}, similar to Fig.~\ref{fig: 2D polar model fig depth}a.  }
\label{fig: 3D fig}
\end{figure}

Figure~\ref{fig: 3D fig} illustrates an example of how $\psi$ extends downward and intersects with the MWR channels for $N = 3 \times 10^{-3} \,\mathrm{s}^{-1}$, the lower bound estimated in \citeA{Lee2021}. Channel~6, sensitive to pressures around $0.7$ bar, lies slightly above the eigenfunctions' range (which starts near 1 bar). We assume that Channel~6 will display a $\psi$ field similar to that of Channel~5. The JIRAM background photograph \cite{adriani2018} represents the cloud deck at approximately this pressure level. The downward extension of $\psi$ below Channel~4 depends on the excitation of mode~1. It is important to note that the MWR measures brightness temperature, which is only indirectly related to $\psi$. Predicting MWR observations requires a layer of translation, involving solutions for vertical velocity and the subsequent diffusion-advection of ammonia, given the background ammonia distribution \cite{Duer2021}.

\section*{Discussion}

In this study, we examined how the vertical extent of polar cyclones influences their vorticity-driven motion (Fig.~\ref{fig: beta-drift with Bu}). Our analysis demonstrated that shallower cyclones undergo a weaker $\beta$-drift, which explains the order-of-magnitude discrepancy between the observed westward drift of the south-polar cyclones and the much faster rates anticipated by an idealized, infinitely deep model (\citeA{Gavriel2023}, their Fig.~2).

We then employed a 2D polar model initialized with identical cyclones at their observed locations (during PJ4). This model provides a dynamical constraint on the deformation radius, $L_d$ (Fig.~\ref{fig: 2D polar model fig depth}), which encodes information about the cyclones' vertical structure. We found that both excessively large and excessively small values of $L_d$ produce westward drifts that deviate strongly from observations. A sampling algorithm estimates $L_d \sim 226 \,(\mathrm{km})$ at the north pole and $\sim 364 \,(\mathrm{km})$ at the south pole. While the posterior distributions of $L_d$ at the two poles do overlap, it remains an open question whether genuine differences in $L_d$ exist between the poles, especially given Jupiter's negligible obliquity and the resulting expected hemispheric symmetry. However, the indicated difference in $L_d$ between the poles could have important implications for the characteristics of the polar cyclones. 
Since $L_d$ controls the spatial scale of vortices in geostrophic turbulence \citep{Polvani1994}, the possibly smaller $L_d$ at the north pole may explain the greater number of smaller cyclones observed at the north pole compared to the south.

Using these estimates of $L_d$, we sought consistency between the 2D QG framework (Fig.~\ref{fig: 2D polar model fig depth}) and a 3D QG model that incorporates density variations with depth (Eq.~\ref{eq: 3D QG equation, depth}). This required solving the eigenvalue problem in Eq.~\ref{eq: eigenvalue problem, depth}, in which the eigenvalues yield $L_d$ for different vertical modes, and the numerically determined eigenfunctions describe the cyclone's vertical structure relative to the cloud deck. In our idealized approach, we assumed a constant Brunt-V\"ais\"al\"a frequency, $N$, and a single dominant vertical mode, both of which can introduce uncertainty. 

We emphasize that the eigenvalue-derived $L_d$ represents a vertically integrated quantity that encapsulates, as accurately as possible, the effects of vertical stretching on 2D dynamics when density varies significantly with depth and the atmosphere is stably stratified. While the linear dynamics at the top layer ($z = 0$) of the 3D-QG model (Eq.~\ref{eq: 3D QG equation, depth}) and the 2D-QG model (Eq.~\ref{eq: 2D QG equation, depth}) are identical---for example, in a Rossby wave solution \cite{vallis2017atmospheric}---nonlinearities and multi-modal dynamics introduce deviations between the models (see SI). Quantifying this error is worthwhile but requires full 3D simulations at the pole with sufficient resolution, which is left for future work.

The solutions (Fig.~\ref{fig: Depth eigenprob}) suggest that for $N \approx 10^{-2} \,\mathrm{s}^{-1}$, cyclone depths of $10-75 \,$ km at the north pole and $20-90 \,$ km at the south pole are plausible, depending on the mode. If $N \approx 3\times10^{-3} \,\mathrm{s}^{-1}$, the estimated cyclone depths increase substantially, ranging $30-220$ km at the north pole and $60-330 \,$ km at the south pole. As a comparison, the GRS has been estimated to be approximately $300 \pm 100$~km deep based on gravity measurements \cite{Parisi2021}, and its signature has been observed down to MWR channel 1, corresponding to depths of $O(300\,\mathrm{km}$) \cite{Bolton2021}, consistent with both experimental and numerical modeling results \cite{lemasquerier2020remote}. Although the GRS is a distinct low-latitude anticyclone, these estimates suggest that the lower-N ($\sim3\times10^{-3}\,\rm{s}^{-1}$) regime considered here may provide a more realistic representation of Jupiter’s deep vortices.

These results (Fig.~\ref{fig: Depth eigenprob}, \ref{fig: 3D fig}) also establish a framework for interpreting forthcoming measurements from Juno's MWR at Jupiter's north pole \cite{Orton2024epsc}. Once these measurements become available, our model can be used to connect each channel's footprint to the underlying vertical structure and static stability $N$. In a subsequent analysis, one can investigate how the cyclonic circulations drive vertical transport and ammonia redistribution, following an approach akin to that used in the study of Jupiter's Ferrel-cell circulation \cite{Duer2021}.

In summary, our work presents a novel framework connecting observed cloud-level motions to the deeper structure of Jovian polar cyclones. By linking the deformation radius, derived from $\beta$-drift constraints, to vertical modes in a 3D-QG framework, we gain new insights into how these cyclones extend below the visible cloud layer and how their internal structure influences their horizontal drift. These constraints on deformation radius, static stability, and vertical profile can guide future studies aimed at modeling the energy fluxes responsible for the formation and maintenance of these cyclones against dissipation.

\section*{Open Research}
No new data sets were generated during the current study. The data
analyzed in this study were published by Ref.\cite{mura2022five_years} (DOI: \url{https://doi.org/10.1029/2022JE007241}), as cited in the
text. The simulations were run with the Dedalus solver \cite{burns2020dedalus} (DOI: \url{https://doi.org/10.1103/PhysRevResearch.2.023068}).

\subsection*{Acknowledgments}
This research has been supported by the Minerva Foundation with funding
from the Federal German Ministry for Education and Research and the
Helen Kimmel Center for Planetary Science at the Weizmann Institute
of Science. We thank Rei Chemke, Keren Duer, Eli Galanti and Or Hadas for insightful conversations and helpful feedback.

\subsection*{Competing Interests}

Authors declare that they have no competing interests.

\FloatBarrier
\bibliographystyle{agu.bst} \bibliography{NimrodObsBib}

\renewcommand{\thefigure}{{\arabic{figure}}}
\setcounter{figure}{0}     
\renewcommand{\thetable}{S\arabic{table}} 
\renewcommand{\thefigure}{S\arabic{figure}}
\renewcommand{\theHtable}{Supplement.\thetable}
\renewcommand{\theHfigure}{Supplement.\thefigure}
\appendix
\onecolumn

\section*{Intuition for the Stretching Term from a Scaling Argument}

\begin{figure}[t]
    \centering
    \includegraphics[width=0.7\columnwidth]{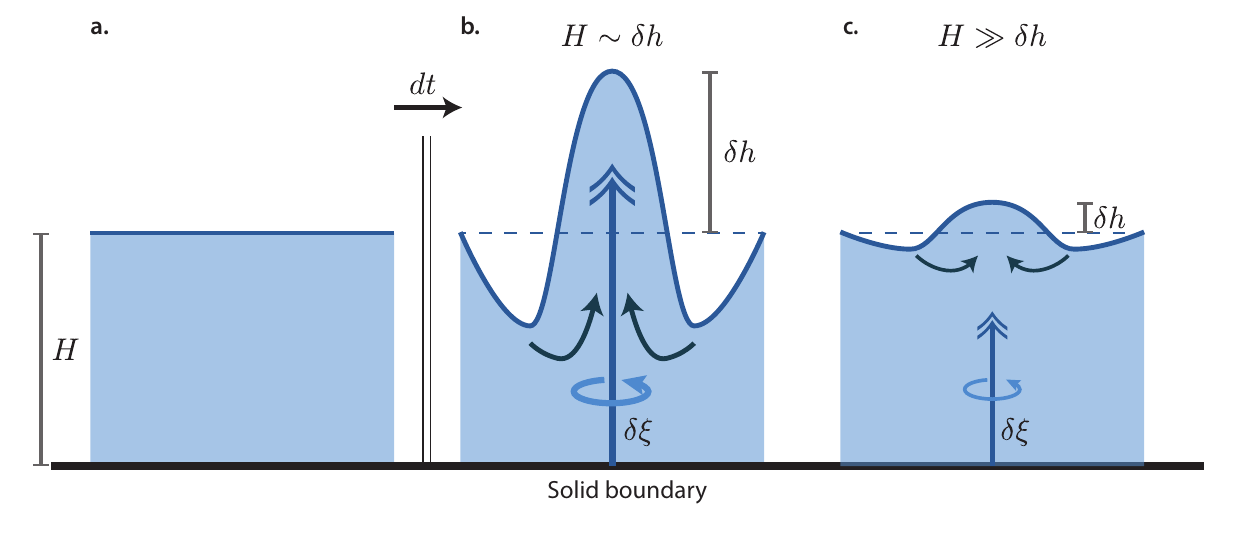}
    \caption{Illustration of how stretching affects shallow-water flow. (a) A tranquil, rotating fluid layer with a solid lower boundary, a free surface, and mean depth $H$. (b) A significant perturbation induces strong horizontal convergence, producing a depth anomaly $\delta h$ comparable to $H$ and resulting in a large increase in relative vorticity. (c) Similar convergence occurs in a much deeper layer ($H \gg \delta h$), leading to a negligible relative vorticity anomaly due to the weaker effect of stretching.\label{fig:stretching illustration}}
\end{figure}

An alternative form of the barotropic potential vorticity (PV) equation is given by \cite{vallis2017atmospheric}:
\begin{equation}
    \frac{\mathrm{D}}{\mathrm{D}t}\left(\xi + \beta y - \frac{f_0}{H} \delta h \right) = 0,
\end{equation}
where $\delta  h$ is the anomaly of the layer depth (Fig.~\ref{fig:stretching illustration}). Intuition for the stretching mechanism can be developed by considering how a small perturbation in depth ($\delta h$) leads to a change in relative vorticity ($\delta\xi$). A scaling argument gives:
\begin{equation}
    \delta\xi = \frac{f_0}{H} \delta h.
\end{equation}
Thus, an increase in layer depth, typically due to horizontal convergence of the flow ($-\nabla \cdot \mathbf{u}$), which conserves mass, generates positive relative vorticity. This vorticity increase is proportional to the background planetary vorticity being concentrated, and inversely proportional to the mean layer depth. Therefore, shallower layers exhibit a stronger stretching effect for a given height anomaly. Conversely, in sufficiently deep layers ($H \gg \delta h$; Fig.~\ref{fig:stretching illustration}c), the influence of stretching on the dynamics becomes negligible.

\section{Developing a Model Following a Cyclone That Includes the Effect of Stretching \label{sec: Depth-methods smith90 model formulation}}

This section provides additional details for the model used to generate Fig.~1. We first derive the governing vorticity equation in a moving reference frame centered on the vortex, then describe the numerical setup and present key aspects of the simulation results.

\subsection*{Deriving the Model Equation}
In this subsection, we derive the vorticity equation with moving coordinates originating in the vortex's center. The model, inspired by \cite{smith1991analytic}, assumes a $\beta$ plane, and one layer of fluid with a constant density.

We begin with the single-layer shallow-water (SW) quasi-geostrophic (QG) vorticity equation on a $\beta$ plane \cite{vallis2017atmospheric}:
\begin{equation}
\frac{D}{Dt}\left(\xi+\beta y-\frac{1}{L_{d}^{2}}\psi\right)=0,\label{eq: Barotropic PV equation}
\end{equation}
where
\begin{equation}
\frac{D}{Dt}=\left(\frac{\partial}{\partial t}+\mathbf{u}\cdot\nabla\right),
\end{equation}
$\mathbf{u}$ is the velocity vector, $\xi = \nabla \times \mathbf{u}$ is the relative vorticity, $\beta$ is the constant meridional gradient of the planetary vorticity ($f$) or any prescribed background vorticity gradient, $L_d=\sqrt{gH}/f_0$ is the deformation radius, $g$ is the gravitational acceleration, $H$ is the layer's depth, and $f_0$ is the reference background rotation rate.

\paragraph{Scaling the Equation}
We non-dimensionalize using the following scales:
\begin{equation}
\begin{array}{c}
\left(x,y\right)=R\left(\hat{x},\hat{y}\right),\;\;\;t=\frac{R}{V}\hat{t},\\
\mathbf{u}=V\hat{\mathbf{u}},\;\;\;\xi=\frac{V}{R}\hat{\xi},\;\;\;\psi=VR\hat{\psi},
\end{array}\label{eq: scales}
\end{equation}
where $R$ and $V$ are the characteristic length and velocity scales of the vortex. In these non-dimensional variables, the vorticity equation becomes
\begin{equation}
\frac{D}{Dt}\left(\xi+\hat{\beta}y\right)-\frac{1}{{\rm Bu}}\frac{\partial\psi}{\partial t}=0,\label{eq: Barotropic PV nondim}
\end{equation}
where
\begin{equation}
\hat{\beta}\equiv\frac{\beta R^{2}}{V},\;\;\;{\rm Bu}\equiv\frac{L_{d}^{2}}{R^{2}}=\frac{gH}{R^{2}f_{0}^{2}},\label{eq: scales-1}
\end{equation}
and $\hat{\beta}$ is a small number, representing the magnitude of $\beta$-drift. This non-dimensional number can be intuitively understood as a ratio between the change in vorticity across the vortex ($\delta f=\beta R$) and the vortex's relative vorticity scale ($V/R$).

\paragraph{Moving to a Coordinate System That Drifts With the Vortex}
We shift to a moving coordinate system $\left(x_\beta, y_\beta\right)$ such that
\begin{equation}
\begin{array}{c}
x=x_{\beta}+\int_{0}^{t}\hat{\beta}u_{\beta}dt,\\
y=y_{\beta}+\int_{0}^{t}\hat{\beta}v_{\beta}dt,
\end{array}\label{eq: scales-2}
\end{equation}
where $\hat{\beta}\mathbf{u}_{\beta}=\hat{\beta}\left(u_{\beta},v_{\beta}\right)$ is the beta-drift speed, changing only with time, and $\left(x_{\beta},y_{\beta}\right)$ is a new coordinate system that moves with $\mathbf{u}_{\beta}$. By the chain rule (with subscript 0 for the stationary frame):
\begin{equation}
\frac{\partial \psi}{\partial t_0}= \frac{\partial \psi}{\partial t}-\hat{\beta}u_{\beta}\frac{\partial\psi}{\partial x_{\beta}}-\hat{\beta}v_{\beta}\frac{\partial\psi}{\partial y_{\beta}},
\end{equation}
so the material derivative becomes
\begin{equation}
\frac{D_0}{Dt}=\frac{\partial_\beta}{\partial t}+\left(\mathbf{u}-\hat{\beta}\mathbf{u}_{\beta}\right)\cdot\nabla_{\beta},
\end{equation}
where $\nabla_\beta$ is a gradient in the moving coordinates. In the moving coordinates, the vorticity equation becomes
\begin{equation}
\biggl[\frac{\partial}{\partial t} + 
\left(\mathbf{u} - \hat{\beta}\,\mathbf{u}_{\beta}\right)\cdot\nabla_{\beta}\biggr]
\Bigl(\,\xi + \hat{\beta}\bigl[y + \hat{\beta}\!\int_{0}^{t} v_{\beta}\,dt\bigr]\Bigr)
\;-\;
\frac{1}{\mathrm{Bu}}\biggl(\frac{\partial \psi}{\partial t} - \hat{\beta}\,\mathbf{u}_{\beta}\cdot\nabla_{\beta}\psi\biggr) 
\;=\; 0.\label{eq: Barotropic PV nondim-1}
\end{equation}
We then decompose the fields into a fixed (vortex) component and a perturbation (or generated) component:
\begin{equation}
f=f_{\rm v}(x_{\beta},y_{\beta})+\hat{\beta}f_{\rm g}(x_{\beta},y_{\beta},t).
\end{equation}
After dividing by $\hat{\beta}$ and retaining terms at leading order, we obtain
\begin{equation}
\frac{\partial}{\partial t}\left(\xi_{{\rm g}}-\frac{\psi_{{\rm g}}}{{\rm Bu}}\right)=-v_{{\rm v}}-\mathbf{u}_{{\rm v}}\cdot\nabla\xi_{{\rm g}}-\left(\mathbf{u}_{{\rm g}}-\mathbf{u}_{\beta}\right)\cdot\nabla\xi_{{\rm v}}-\frac{1}{{\rm Bu}}\mathbf{u}_{\beta}\cdot\nabla\psi_{{\rm v}}+O(\hat{\beta}), \label{eq: Barotropic PV nondim moving coordinates}
\end{equation}
while noting that due to the symmetry of the constant vortex profile, advective terms $\mathbf{u}_{\rm v}\cdot\nabla (\psi_{\rm v}$ or $\xi_{\rm v})$ are zero. This equation represents a potential vorticity equation for the evolution of a velocity field generated by $\beta$-drift in a moving frame. The velocity vector, $\mathbf{u}_{\beta}\approx \mathbf{u}_{\mathrm{g}}(0,0,t)$ is the generated velocity at the vortex center, and $\mathbf{u}_{\mathrm{v}}, \xi_{\mathrm{v}}, \psi_{\mathrm{v}}$ are determined by the chosen vortex profile. 

The version of Eq.~\ref{eq: Barotropic PV nondim moving coordinates} presented in the main text (Eq.~1) is simplified for focusing the reader's attention on the important points. It unscales back the "$\rm{g}$" variables by $\hat\beta$, so as to directly see the dependence of $\beta$-drift on $\hat\beta$, and assumes that the terms $\left(\left(\mathbf{u}_{{\rm g}}-\mathbf{u}_{\beta}\right)\cdot\nabla\xi_{{\rm v}},\;\;-\frac{1}{{\rm Bu}}\mathbf{u}_{\beta}\cdot\nabla\psi_{{\rm v}}\right)$ are relatively small. This assumption can be intuitively understood at the beginning of the motion, where $\mathbf{u}_{{\rm g}}=\mathbf{u}_{\beta}=0$. Nonetheless, the simulations (Fig.~1) use the full version (Eq.~\ref{eq: Barotropic PV nondim moving coordinates}).

\subsection*{Numerical Model Setup}

For solving Eq.~\ref{eq: Barotropic PV nondim moving coordinates} we use the Dedalus PDE solver \cite{burns2020dedalus} with double periodic bases. To enforce a concentric boundary, we set the (scaled) domain size as $(x,y)\in\{-20,20\}$ and add a relaxation term of the form 
\begin{equation}
    \frac{\xi_g}{\tau_{\rm{d}}}  G(r,R_{\rm{d}},A_{\rm{d}}),
\end{equation}
where $\tau_{\rm{d}}=0.1\pi$ is the relaxation time scale, 
\begin{equation}
    G=\frac{1}{2}\left(1-\tanh{\left(A_{\rm{d}}\left(R_{\rm{d}}-r\right)\right)}\right)
    \label{eq: activation function}
\end{equation}
is a continuous activation function, $r=\sqrt{x^2+y^2}$ is the radius from the pole, $A_{\rm{d}}=20$ is the sharpness of the transition to the relaxation zone, and $R_{\rm{d}}=15$ is the activation radius. We also added a viscosity term of the form $\rm{Ek}\nabla^4\psi_{\rm{g}}$, with $\rm{Ek}=0.01$, for stability. A resolution of $96\times96$ grid points was sufficient for the parameter sweep (Fig.~1).  

The vortex profile (compare to Eq.~2 in the main text, here in scaled variables) is taken as \cite{chan1987analytical}
\begin{equation}
\xi_{\rm v} 
= \left( 2 -r^b \right) 
e^{\frac{1}{b}\left(1 - r^b\right)},
\label{eq: chan87 vorticity profile SI}
\end{equation}
where $b$ controls how sharply the tangential velocity changes with radius. From $\xi_{\mathrm{v}}$, we obtain $\psi_{\mathrm{v}}$ by solving the Poisson equation $\nabla^2 \psi_{\mathrm{v}} = \xi_{\mathrm{v}}$.

\begin{figure}[t]
    \centering
    \includegraphics[width=1\columnwidth]{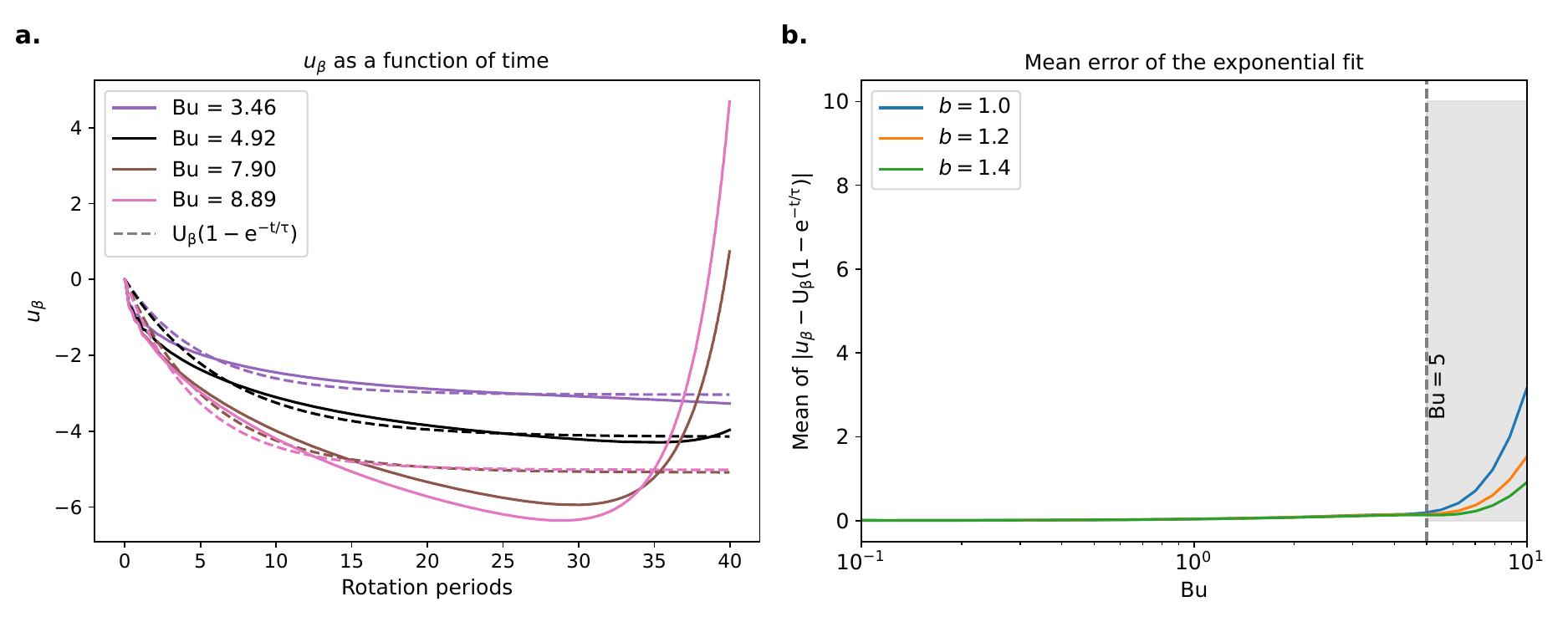}
    \caption{\textbf{Analysis of the exponential fit, $\rm{U}_\beta(1-e^{-t/\tau})$, adequacy.} (a) The same as Fig.~1a, but for different $\rm Bu$ values, showing poor exponential fit for $\rm{Bu}=(7.9,8.89)$. (b) The mean integrated absolute error between the  $u_\beta$ and the exponential fit. It can be seen that for $\rm Bu\gtrsim5$ the exponential parametrization is not adequate for capturing $u_\beta(t)$.  
    \label{fig: Exp error}}
\end{figure}

In Fig.~1(a-b), we plot the resulting $(u_\beta,v_\beta)$ with $b=1.2$ and four values of $\rm{Bu}$. We find that $u_\beta$ behaves like an exponential relaxation function, $u_\beta\sim\rm{U}_\beta\left(1-e^{-t/\tau}\right)$, where the amplitude $\rm{U}_\beta$ and timescale $\tau$ minimize the root-mean-square error (RMSE) relative to the simulated $u_\beta$. This parameterization allows for a simple presentation of the behavior of $u_\beta$ as a function of $\rm Bu$ (Fig.~1c). We observe that this parameterization becomes poorly fitted for values of $\rm Bu \gtrsim 5$, as illustrated in Fig.~\ref{fig: Exp error}.

As seen in Fig.~1a–b, $\mathbf{u}_\beta$ exhibits oscillations at the beginning of the motion. This likely results from the following mechanism: the vortex core rotates significantly faster than its outer regions, and $\beta$-gyres initially develop within the core, aligned in the east–west direction. These gyres begin to rotate with the core, generating the observed wiggling in the $\beta$-drift velocity. The oscillations gradually subside as the core homogenizes the background vorticity gradient and as the gyres propagate outward from the core. This behavior is illustrated in Movie~S1, which shows the simulation corresponding to Fig.~1e. A Dedalus model script for solving Eqs.~\ref{eq: Barotropic PV nondim moving coordinates}–\ref{eq: chan87 vorticity profile SI} is available on \href{https://github.com/nimrodgav/Dynamical-constraints-on-the-vertical-structure-of-Jupiter-s-polar-cyclones}{GitHub} (\texttt{beta\_plane\_moving\_frame\_vortex.py}).

\section{The Polar QG Model}

In this section, we present the details of the polar QG model and the statistics underlying the results shown in Fig.~2. We solve the 2D-QG equation \cite{vallis2017atmospheric}, 
\begin{equation}
\frac{D}{Dt}
\left(\nabla^2\psi 
+ f
- \frac{1}{L_{d}^2}\psi 
\right) = 0,
\label{eq: 2D QG equation, SI}
\end{equation}
where 
\begin{equation}
    f=2\Omega\cos\left(\frac{\sqrt{x^2+y^2}}{R_{\rm{Jo}}}\right), \label{eq: Coriolis sphere}
\end{equation}
and $\Omega=1.759\times10^{-4}$ is Jupiter's rotation rate. While Eq.~\ref{eq: Coriolis sphere} assumes a spherical planet, Jupiter’s oblateness necessitates a correction. We incorporate this by using the osculating radius, $R_{\rm Jo} = R_{\rm Je}^2 / R_{\rm Jp} = 76{,}452$ km, where $R_{\rm Je} = 71{,}492$ km and $R_{\rm Jp} = 66{,}854$ km are the equatorial and polar radii, respectively \cite{hyder2022}. This adjustment ensures the Coriolis term reflects the correct curvature near the polar regions. We note that, for translating between latitude–longitude and Cartesian ($x$–$y$) coordinates of the cyclones, we use Jupiter’s polar radius, which is more appropriate given the local geometry near the poles.

In practice, our model solves the non-dimensional form of Eq.~3:
\begin{equation}
\frac{D}{Dt}
\left(\nabla^2\psi 
+ \hat{f} 
- \frac{1}{\rm Bu}\psi 
\right) = \rm{Ek}\nabla^4\psi,
\label{eq: 2D QG equation, scaled}
\end{equation}
where lengths are scaled by $L=1{,}000$ km and velocities by $U=50$ m\,s$^{-1}$. Thus, time is scaled by $T=L/U$. The non-dimensional variables are $\hat{\psi} = \tfrac{1}{LU}\,\psi$ and $\hat{f} = \tfrac{L}{U} \, G(r,R_{\mathrm{d}},A_{\mathrm{d}})\, f$. We set the numerical viscosity to $\mathrm{Ek} = 10^{-5}$, and the Burger number is $\mathrm{Bu} = (L_{d}/L)^2$, where $L_d$ is an input parameter.

 We solve the model using Dedalus \cite{burns2020dedalus} on a doubly periodic Fourier domain. The initial conditions consist of $N_{\mathrm{cyc}}$ cyclones (9 for the north pole and 6 for the south pole) of the form (in dimensional variables):
\begin{equation}
\xi _{(t=0)}
= \sum_{n=1}^{N_{\rm{cyc}}} \frac{V}{R} \left( 2 - \left(\frac{r_n}{R}\right)^b \right) 
\exp\left[\frac{1}{b}\left(1 - \left(\frac{r_n}{R}\right)^b\right)\right],
\label{eq: chan87 vorticity profile mult}
\end{equation}
where $r_n = \sqrt{(x - x_n)^2 + (y - y_n)^2}$ measures the distance from the center of cyclone $n$. The cyclone centers $\{(x_n,y_n)\}$ are taken from Juno PJ4 observations \cite{mura2022five_years}, with the original planetocentric latitude coordinates converted to Cartesian positions using Jupiter’s polar radius, $R_{\rm Jp}$. We obtain the initial streamfunction by solving $\nabla^2 \psi = \xi_{(t=0)}$ prior to each run.

To mitigate boundary effects from the rectangular domain, we use the "trap" method \cite{Siegelman2022}. Specifically, we multiply the planetary PV by an activation function (Eq.~\ref{eq: activation function}) which smoothly transitions from $\approx 0$ outside $r>R_{\mathrm{d}}$ to $\approx \tfrac{L}{U}f$ inside $r<R_{\mathrm{d}}$. In this study, we use $A_{\mathrm{d}}=-20$ (negative values switch the domain of activation to within the circle), and choose $R_{\mathrm{d}}$ to be 97\% of the total domain size. This procedure effectively "disconnects" the dynamics of interest (inside $R_{\mathrm{d}}$) from the spurious influence of the periodic boundaries (outside $R_{\mathrm{d}}$).  Throughout each simulation, we track the centers of the cyclones in time and maintain their individual identities, thereby obtaining a record of their trajectories and possible mergers.
A Dedalus model script for solving Eqs.~\ref{eq: 2D QG equation, scaled}–\ref{eq: chan87 vorticity profile mult} is available on \href{https://github.com/nimrodgav/Dynamical-constraints-on-the-vertical-structure-of-Jupiter-s-polar-cyclones}{GitHub} (\texttt{polar\_cyclones.py}) along with the tracking analysis script (\texttt{polar\_cyclones\_cost.py}).

\subsection*{Sensitivity to Resolution and Domain Size}
Because we sample the model repeatedly, we first identified the smallest resolution and domain size that would produce acceptably accurate cyclone trajectories. As a benchmark, we ran a high-resolution model ($360 \times 360$ grid points) with a large domain ($36{,}000$ km on each side). We initialized nine identical cyclones at the north polar PJ4 positions $\bigl(L_d=348\;\mathrm{km}, R=867\;\mathrm{km}, V=86.13\;\mathrm{m\,s}^{-1}, b=1.51\bigr)$, and integrated for $2{,}500$ days.

\begin{figure}[t!] 
\begin{centering} 
\makebox[\textwidth][c]
{\includegraphics[width=1\textwidth]{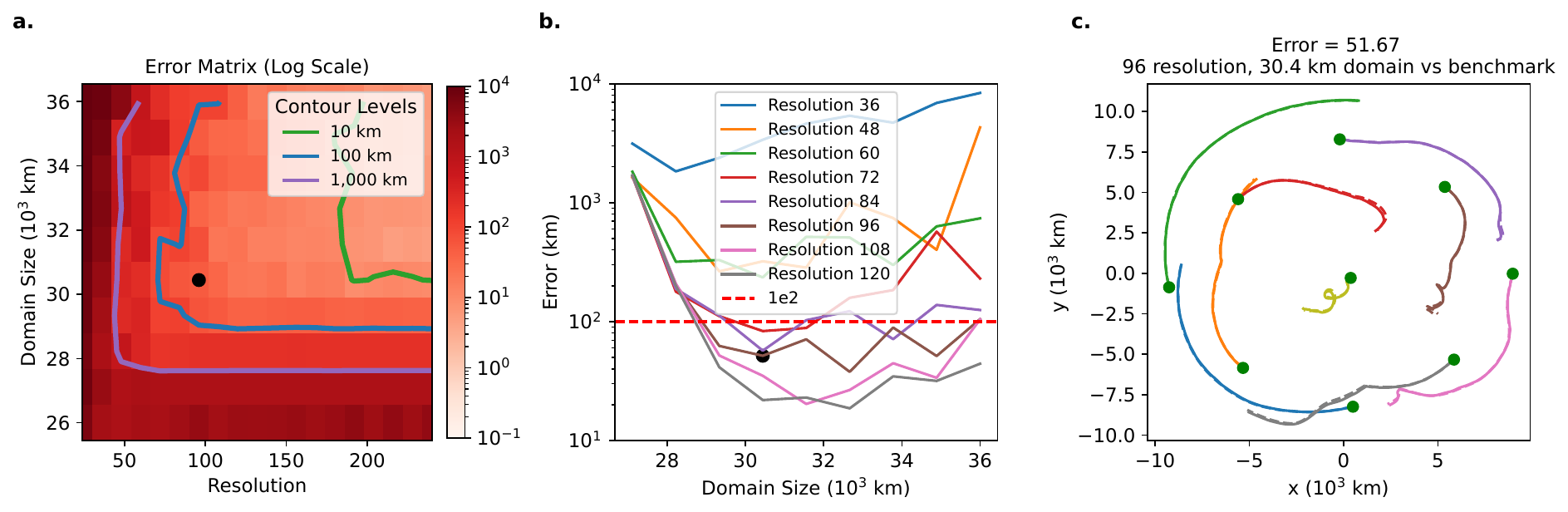}}
\par\end{centering} 
\caption{\textbf{Analyzing the required resolution and domain size for sampling the model}. (a) Contours of RMSE (in km) between each model's cyclone trajectories and the benchmark, as functions of resolution (abscissa) and domain size (ordinate). The blue contour highlights the 100 km threshold for acceptable error. The black dot is the chosen setting. (b) RMSE dependence on domain size for selected resolutions. (c) Benchmark trajectories (solid lines) compared with those at the chosen $96\times96$ resolution and a $30{,}400$ km domain (dashed lines).
\label{fig: polar QG determine resolution}} 
\end{figure}

Next, we reran the model at coarser resolutions and smaller domain sizes. For each simulation, we computed the RMSE between the resulting cyclone trajectories and those from the benchmark (Fig.~\ref{fig: polar QG determine resolution}). We are primarily interested in the average westward drift of these cyclones, so we allowed a maximum mean trajectory error of 100 km (blue contour in Fig.~\ref{fig: polar QG determine resolution}a). We found that a $96\times96$ grid with a domain size of $30{,}400$ km satisfies this requirement. Figure~\ref{fig: polar QG determine resolution}b,c illustrates that the associated trajectories agree well with the benchmark. We therefore adopted these numerical settings for the sampling runs used to generate Fig.~2.

\subsection*{The Sampling Algorithm}
To constrain $L_d$ (the deformation radius) for Jupiter's north and south poles using the observed westward drift rates, we apply Bayesian nested sampling via the \texttt{Ultranest} library \cite{buchner2021ultranest}. This algorithm explores parameter space efficiently, according to a likelihood function, and returns a posterior distribution of unknown parameters.
 
 While noting that the model is idealized, and that we assume equivalent cyclones in order to drastically reduce the number of unknown variables, we are left with the 3 cyclone variables ($R,V,b$) and $L_d$ as our 4 unknown variables. We assign the following normal prior distributions to these parameters, denoted $\mathcal{N}(\mu,\sigma^2)$, where $\mu$ is the mean and $\sigma$ is the standard deviation of the distribution:
\begin{equation}
\begin{aligned}
    L_d &\sim \mathcal{N}(400, 150^2) \quad \rm{(\rm{km})}, \\
    b &\sim \mathcal{N}(1.2, 0.2^2), \\
    U &\sim \mathcal{N}(80, 15^2) \quad \rm{(\rm{m/s})}, \\
    R &\sim \mathcal{N}(900, 150^2) \quad \rm{(\rm{km})}.
\end{aligned}
\end{equation}

We construct the likelihood function with two components. The first term,
\begin{equation}
    C_1=-\frac{1}{2}\left(\dfrac{\overline{\mathrm{WW}}_{\rm O} - \overline{\mathrm{WW}}_{\rm M}}{5\times10^{-3}\,\mathrm{m\,s}^{-1}} \right)^2
\end{equation}
penalizes deviations between the observed mean westward drift $\overline{\mathrm{WW}}_{\rm O}$ and the model's mean drift $\overline{\mathrm{WW}}_{\rm M}$. The observed drift is averaged over five years and $N_{\mathrm{cyc}}$ cyclones, while the model drift is averaged over five years after a two-year spinup. The drifts are calculated like in \citeA{Gavriel2023}, without taking differences between the observed cyclones into account. 

The second term,
\begin{equation}
    C_2 = -\frac{1}{2}\left(\dfrac{N_{\rm mergers}}{2}\right)^2,
\end{equation}
penalizes runs producing cyclones that merge. We simulate seven years of model time in snapshots spaced 53 days apart (approximately the period of Juno's orbits). Each "missing" cyclone (due to merger) in a snapshot contributes to $N_{\mathrm{mergers}}$. Models with more frequent or earlier mergers thus have increasingly negative $C_2$. 

\begin{figure}[t!]
    \centering
    \includegraphics[width=1\linewidth]{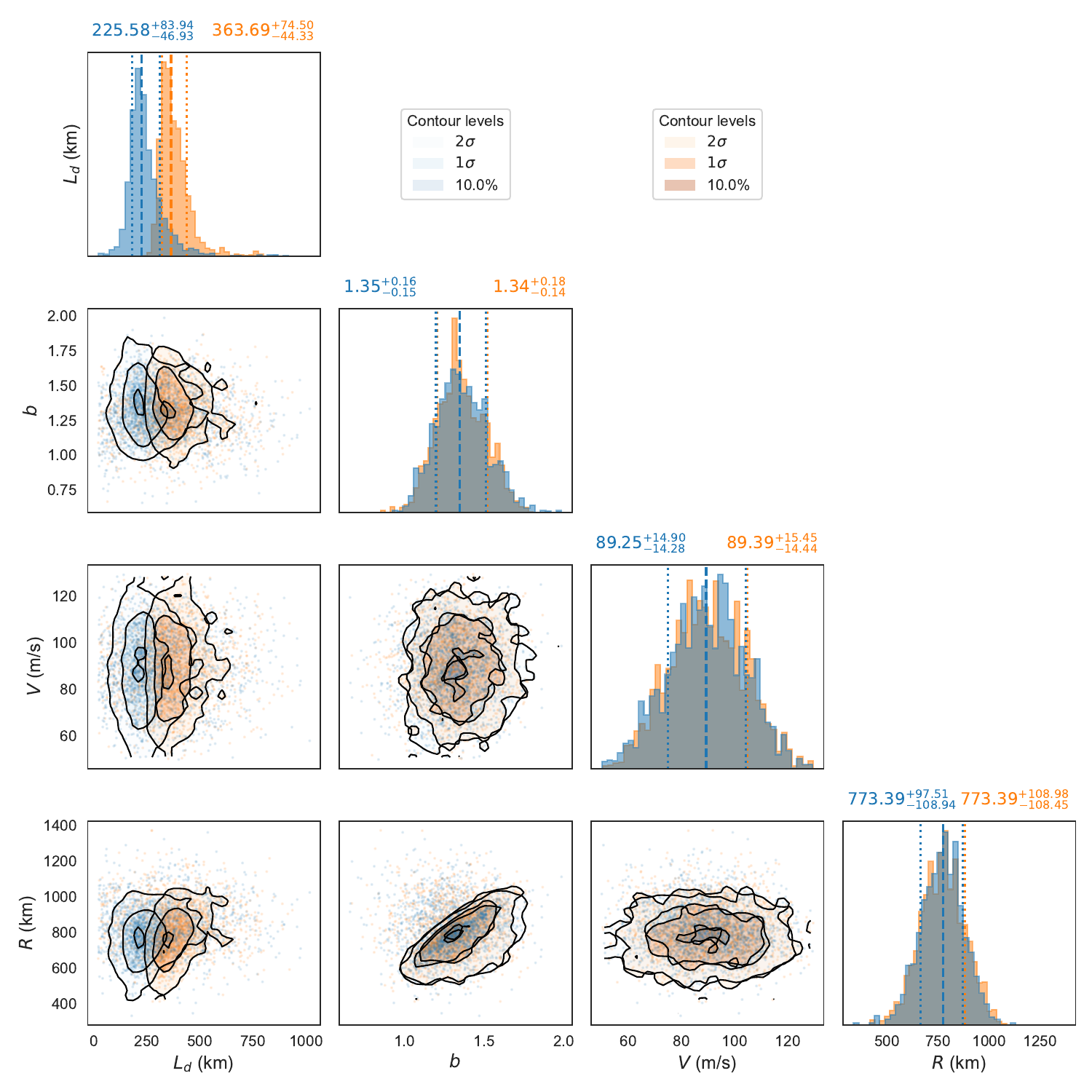}
    \caption{Joint and marginal posterior distributions of the model parameters for the North Pole (blue) and South Pole (orange).
    Diagonal panels are the same as Fig.~2k-n; off-diagonal panels display the joint densities with filled contours at the $10 \%$, $68.27 \%\, (1\sigma)$ and $95.45 \%\, (2\sigma)$ levels overlaid with individual samples (points).}
    \label{fig:corner_plot}
\end{figure}

These two components combine additively:
\begin{equation}
C=C_1 + C_2,    
\end{equation}
and we use this total likelihood to guide the nested-sampling algorithm. Each pole (north and south) is sampled with $15{,}000$ allowed evaluations, yielding the posterior distributions presented in Fig.~2k-n. The joint posterior distributions, shown as corner plots in Fig.~\ref{fig:corner_plot}, reveal that among all parameter pairs, only $b$ and $R$ exhibit a simple linear correlation, while the remaining parameters display more complex relationships.

\section{The Brunt-V\"ais\"al\"a  Frequency of Jupiter's Atmosphere}

\begin{figure}[t]
    \centering
    \includegraphics[width=0.9\linewidth]{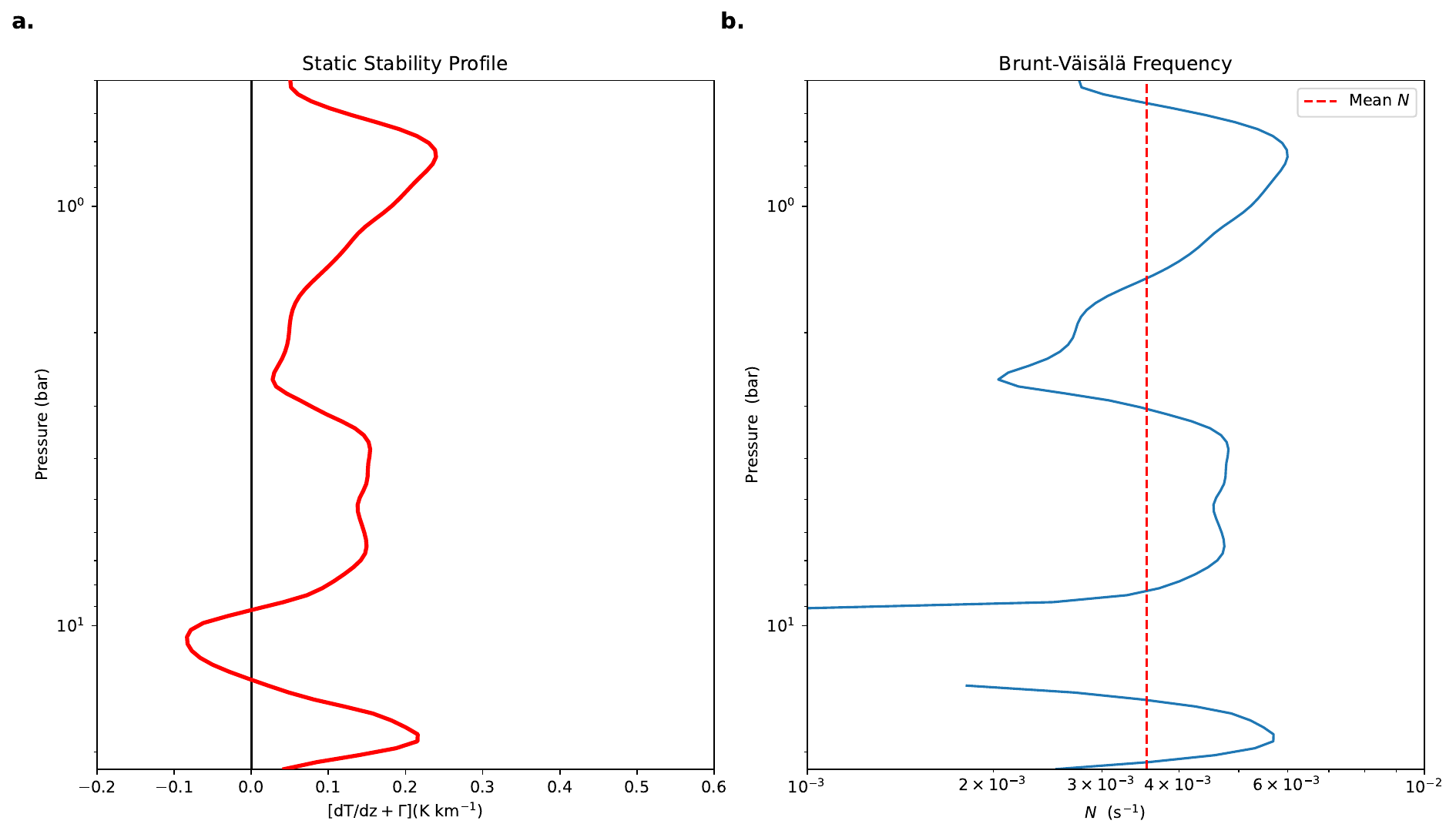}
    \caption{The Brunt V\"ais\"al\"a ($N$) frequency from the Galileo probe decent. (a) The static stability, $S$, digitized from Fig.~7 in \citep{Magalhaes2002}. (b) Calculation of $N$ based on panel (a). The red vertical line indicates the pressure-weighted mean value of $N$ over the range $1 \leq p \leq 22$~bar.}
    \label{fig: Brunt}
\end{figure}

The results shown in Fig.~3 depend on the vertical profile of the Brunt-V\"ais\"al\"a frequency ($N$, in units of s$^{-1}$). The only direct measurement of temperature at pressures greater than 1 bar comes from the Galileo probe, which entered Jupiter’s atmosphere on December 7, 1995, at $6.5^\circ$N and transmitted data down to approximately 22 bar \citep{Seiff1996}. A detailed analysis of the vertical static stability from this measurement was presented by Magalh\~aes et al.(2002) \citep{Magalhaes2002}, and is digitized in Fig.~\ref{fig: Brunt}a.

In their analysis, static stability is expressed as the difference between the vertical temperature gradient and the dry adiabatic lapse rate:
\begin{equation}
S = \frac{\partial T}{\partial z} + \Gamma \left(\mathrm{K}\;\mathrm{km}^{-1}\right).
\end{equation}
We then compute $N$ (Fig.~\ref{fig: Brunt}b) using the relation:
\begin{equation}
N^{2} = \frac{g}{T} \left( \frac{\partial T}{\partial z} + \Gamma \right) = \frac{g}{T} S.
\end{equation}

However, there are several caveats to using this profile in our calculations (Fig.~3). First, the Galileo probe descended at an equatorial latitude ($6.5^\circ$N), which is dynamically distinct from the polar region of interest. Second, the probe entered a known atmospheric “hot spot” \citep{Magalhaes2002}, which may not be representative of Jupiter’s mean atmospheric structure. Third, significant uncertainties affect the measurement itself, including unknown vertical wind velocities during descent and thermal disturbances near the probe’s sensors. These factors may explain the (likely spurious) negative static stability values observed near $10$-bar.

Given these uncertainties, the limited depth range of the data, and to ensure consistency and simplicity in our eigenvalue calculations (Fig.~3), we opt to use a constant value of $N$. Nevertheless, the pressure-weighted average,
\begin{equation}
\langle N \rangle = \frac{\int_{1\rm{bar}}^{22~\rm{bar}} N(p) \, p \, dp}{\int_{1~\rm{bar}}^{22~\rm{bar}} p \, dp} \sim 3.7 \times 10^{-3}\mathrm{s}^{-1},
\end{equation}
agrees well with estimates by Lee et al.(2021) \citep{Lee2021}, inferred from Juno-based ammonia distribution, and is referenced accordingly in this manuscript.

\section{Equivalence Between the 3D and 2D QG Models for a Given Mode}

In this section, we provide a mathematical argument for the equivalence between the 2D-QG model results shown in Fig.~2 and the 3D-QG model decomposed into vertical modes, as presented in Fig.~3.

We begin with the 3D-QG equation \cite{vallis2017atmospheric}:
\begin{equation}
\frac{D_h}{Dt}\left(\nabla_h^{2}\psi+f+\frac{f_{0}^{2}}{\tilde{\rho}}\frac{\partial}{\partial z}\left(\frac{\tilde{\rho}}{N^{2}}\frac{\partial\psi}{\partial z}\right)\right)=0,
\label{eq: 3D QG equation, depth, SI}
\end{equation}
where the subscript '$h$' denotes derivatives taken only in the horizontal directions.

Assuming separability of the streamfunction into horizontal and vertical components, we write,
\begin{equation}
\psi(x,y,z,t)=\sum_n \psi_n(x,y,t)\Phi_n(z),
\label{eq: separability expansion SI}
\end{equation}
where the vertical modes $\Phi_n(z)$ are required to be orthogonal:
\begin{equation}
    \langle \Phi_n, \Phi_m\rangle\equiv\int_{-H}^{0}w\,\Phi_n\Phi_m\;dz =C_{nm} \delta_{nm},
\end{equation}
with $w(z) = \tilde{\rho}(z)$ serving as a weight function, $C_{nm}$ constant coefficients, and $\delta_{nm}$ the Kronecker delta. Such orthogonal modes are given by $\Phi_n$, which are non-trivial solutions of the Sturm-Liouville equation,
\begin{equation}
\frac{f_{0}^{2}}{\tilde{\rho}}\frac{\partial}{\partial z}\left(\frac{\tilde{\rho}}{N^{2}}\frac{\partial\Phi_n}{\partial z}\right)+\Gamma_n\Phi_n=0,
\label{eq: eigenvalue problem, depth SI}
\end{equation}
subject to the boundary conditions:
\begin{equation}
    \partial_z\Phi_n(0)=0,
\qquad
\Phi_n(-H)=0. \label{eq: eigenfunction BCs}
\end{equation}
These solutions are thus eigenfunctions ($\Phi_n$ for mode $n$) of Eqs.~\ref{eq: eigenvalue problem, depth SI}-\ref{eq: eigenfunction BCs}, where the eigenvalues are $\Gamma_n$. 

Substituting the modal decomposition (Eqs.~\ref{eq: separability expansion SI}–\ref{eq: eigenvalue problem, depth SI}) into Eq.~\ref{eq: 3D QG equation, depth, SI} yields
\begin{equation}
\frac{D_h}{Dt}\left(\sum_{n}\Phi_{n}\left(\nabla^2_h\psi_n-\Gamma_{n}\psi_{n}\right)+f\right)=0.
\label{eq: 3D QG equation insert sum of modes}
\end{equation}
Linearizing Eq.~\ref{eq: 3D QG equation insert sum of modes}, and projecting onto a single mode by taking the inner product, <$\Phi_n$, Eq.~\ref{eq: 3D QG equation insert sum of modes}>,  gives  
\begin{equation}
    \frac{\partial}{\partial t}\left(\nabla^2\psi_{n}-\Gamma_{n}\psi_{n}\right)+J(\psi_{n},f)=0,
    \label{eq: 2D-QG per mode}
\end{equation}
where $J(a,b) = a_x b_y - a_y b_x$ denotes the Jacobian.

Thus, in the linear regime, each vertical mode of the 3D-QG system evolves independently according to a two-dimensional QG-like equation. This structure mirrors the dynamics described by the linearized 2D-QG equation:
\begin{equation}
\frac{\partial}{\partial t}\left(\nabla^2\psi - \frac{1}{L_d^2} \psi\right) + J(\psi, f) = 0,
\label{eq: 2D QG equation, SI}
\end{equation}
provided that $L_d = \Gamma_n^{-1/2}$. Therefore, for each vertical mode $n$, the linearized 3D-QG dynamics (Eq.~\ref{eq: 3D QG equation, depth, SI}) projected onto $\Phi_n(z)$ are mathematically equivalent to the 2D-QG dynamics (Eq.~\ref{eq: 2D QG equation, SI}) with a mode-dependent deformation radius $L_d$. We argue that, while nonlinear interactions and the coexistence of multiple modes could lead to deviations from this simple structure, the leading-order dependence on $N$ and $L_d$ captured by this modal decomposition provides a robust framework for interpreting the vertical structures expected to be observed by Juno, as shown in Fig.~3.

\subsection*{Including Non-Linear Terms: Triple Interactions}

Performing a modal projection, <$\Phi_n$, Eq.~\ref{eq: 3D QG equation insert sum of modes}>, without linearization, yields
\begin{equation}
K + \frac{\partial}{\partial t}\bigl(\nabla^2\psi_{n} - \Gamma_{n}\psi_{n}\bigr) + J(\psi_{n},f) = 0,
\label{eq: 2D-QG per mode}
\end{equation}
where
\begin{equation}
K = \sum_{i}\sum_{j}\Bigr(\varepsilon_{ijn}J\left({\psi_{i}},\left({\nabla{\psi_{j}}}+\Gamma_{j}\psi_{j}\right)\right)\Bigr)  ,   
\end{equation}
contains both self-modal and inter-modal advection terms, 
\begin{equation}
\varepsilon_{ijn} = \frac{1}{C_{nn}}\int_{-H}^{0}\Phi_{i}\Phi_{j}\Phi_{n}\,\tilde\rho \,dz,
\label{eq: triple coeff}
\end{equation}
is the triple-interaction coefficient \citep{Smith2001b}, and the integral is symmetric under permutations of $(i,j,n)$.

For our two-mode expansion, using the eigenfunctions from Fig.~4 (with $H=50.19\;$km and $N=3\times10^{-3}\;$s$^{-1}$), the nonzero elements for mode~0 are
\begin{equation}
    \varepsilon_{0ij} = \frac{1}{C_{00}}\int_{-H}^{0}\Phi_{0}\Phi_{i}\Phi_{j}\tilde\rho(z)dz=\left[\begin{array}{ccc}
i\setminus j & 0 & 1\\
0 & 0.797 & 0.178\\
1 & 0.178 & 0.508
\end{array}\right].
\end{equation}
Ideally, to achieve an exact correspondence between the 2D and modal 3D QG equations, we would require
\[
\varepsilon_{0ij} =
\begin{cases}
1 & \text{if } i = j = 0, \\
0 & \text{otherwise}.
\end{cases}
\]
In our case, $\varepsilon_{000} = 0.797$ (quantifying the self-advection of mode~0 potential vorticity by the mode~0 velocity field), while the inter-modal coefficients are smaller. 

Despite this partial modal leakage, the overall structure of the 3D QG equation in modal form retains a strong similarity to the 2D QG system, even when nonlinear interactions are included.

\end{document}